\documentclass[journal=jacsat,manuscript=article,layout=twocolumn]{achemso}

\usepackage{chemformula} 
\usepackage[T1]{fontenc} 
\usepackage{times}
\usepackage{chemformula,array}
\usepackage[version=4]{mhchem}

\usepackage{xcolor}
\usepackage[export]{adjustbox}
\usepackage{multirow}

\author{Hunter J. Windsor}
\affiliation{Department of Chemistry, University of Oxford, Inorganic Chemistry Laboratory, South Parks Road, Oxford OX1 3QR, U.K.}
\author{Guy Greenbaum}
\affiliation{Department of Chemistry, University of Oxford, Inorganic Chemistry Laboratory, South Parks Road, Oxford OX1 3QR, U.K.}
\author{Tristan N. Dolling}
\affiliation{Department of Chemistry, University of Oxford, Inorganic Chemistry Laboratory, South Parks Road, Oxford OX1 3QR, U.K.}
\author{Yevheniia Kholina}
\affiliation{Department of Chemistry, University of Oxford, Inorganic Chemistry Laboratory, South Parks Road, Oxford OX1 3QR, U.K.}
\author{Andrew L. Goodwin}
\affiliation{Department of Chemistry, University of Oxford, Inorganic Chemistry Laboratory, South Parks Road, Oxford OX1 3QR, U.K.}
\email{andrew.goodwin@chem.ox.ac.uk}

\title{Hidden Truchet Architecture in\\ Zinc {$\boldsymbol p$}-Hydroxybenzoate}

\begin{document}

\begin{abstract}
We redetermine the structure of the disordered metal--organic framework Zn(hba) (hba$^{2-}$ = the dianion of 4-hydroxybenzoic acid). Using single-crystal X-ray diffraction measurements, we characterise the structured diffuse scattering that is experimentally observed for this material and which is characteristic of strongly correlated disorder. We use geometric and crystal chemical arguments to propose a general model for correlated disorder in Zn(hba), and then relate this model to a specific realisation of so-called Truchet tilings. Using Monte Carlo simulations, we proceed to show that the model so developed is simultaneously consistent with both the average crystal structure solution described previously, and the structured diffuse scattering reported here. The existence of ordered analogues with different, but related, chemistry suggests scope for control over correlated disorder in this family of metal--organic frameworks. Our study illustrates the potential for a Truchet-tile formalism to help describe and understand more generally the correlated disorder that occurs in framework materials---even amongst those that are chemically and crystallographically dissimilar.
\end{abstract}

\small

\section{Introduction}
The incorporation of correlated disorder within materials is an effective methodology for introducing complexity for applications in information storage,\cite{Crutchfield_1994,Cartwright_2012,Goodwin_2025} neuromorphic computing,\cite{Gartside_2022} and error-correcting codes.\cite{Pretzel_1992} It is the configurational degeneracy inherent to disordered states that is responsible for information content,\cite{Kolmogorov_1968,Lieb_1967} and the presence of correlations then gives the nonlinear responses from which other useful functionalities arise.\cite{Crutchfield_2011} Correlated disorder occurs in various guises amongst most materials classes,\cite{Keen_2015,Simonov_2020} but we have become particularly interested in its phenomenology in metal--organic frameworks (MOFs).\cite{Dixey2019,Meekel2021,Ehrling2021, Griffin_2025} Our interest is grounded in the well-established geometric rules known to govern MOF design---\emph{i.e.}, their modular assembly \emph{via} directed metal--ligand coordination chemistry\cite{Yaghi1998,Robson_2000,James2003,Yaghi_2003}---with the obvious prospect for extension to allow the targeted and rational design of their structures containing specific kinds of correlated disorder.

\begin{figure}[b]
    \includegraphics{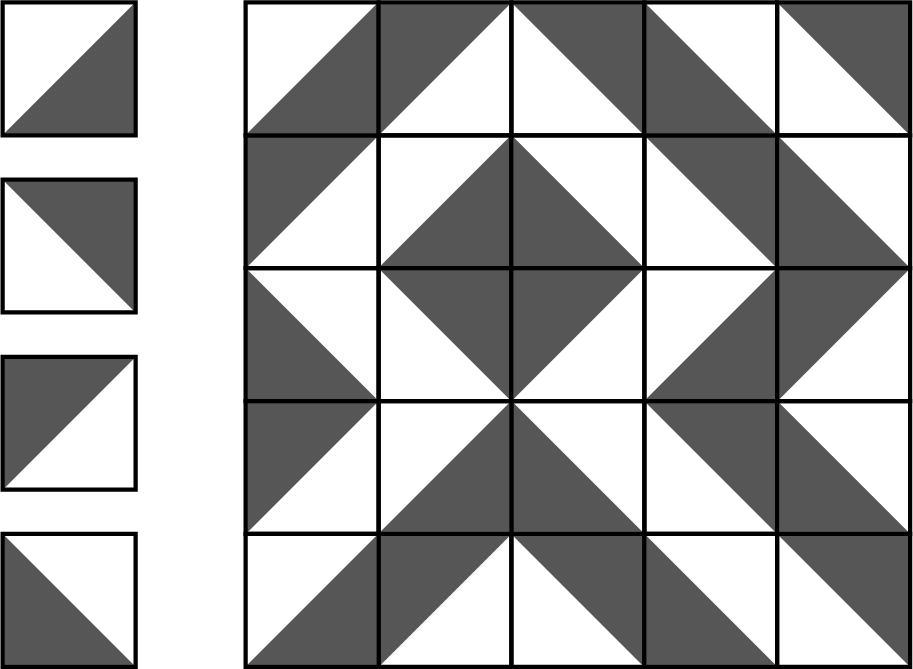}
    \caption{\footnotesize Truchet tilings arise when high-symmetry tiles are decorated to lower their symmetry. Here, the simple diagonal decoration distinguishes four orientations of a square tile. Enforcing matching rules generates complex patterns that are neither ordered nor random.}
    \label{fig1}
\end{figure}

A useful paradigm for understanding correlated disorder in MOFs is that of Truchet tilings [Fig.~\ref{fig1}].\cite{Truchet1704,Meekel2023} These are decorated tilings in which local symmetry-breaking stores information through tile orientation, while matching rules generate correlations between neighbouring tiles. Applied initially in the structural interpretation of [OZn$_4$][1,3-benzenedicarboxylate]$_3$ (TRUMOF-1),\cite{Meekel2023} the concept involves relating the average crystal structure to a space-filling packing of ``node'' and ``linker'' tiles, and then treating the nodes and linkers as (chemically) decorating the tiles to lower their symmetry. There are, by construction, many equivalent orientations for each tile, capturing the varied arrangements of nodes and linkers that occur within a disordered MOF. At the same time, the need to maintain sensible chemical connectivities on a local scale maps to the Truchet-tile matching rules, in turn reflecting the presence and nature of disorder correlations.

We show that this same conceptual methodology is useful beyond the TRUMOF-1 system and allows reinterpretation of a structurally distinct disordered MOF known as Zn(hba) (hba$^{2-}$ = the dianion of 4-hydroxybenzoic acid). The structure of this material was first reported in Ref.~\citenum{White2015}, where the point was made that its X-ray diffraction pattern contained diffuse features. Although not analysed at the time, such features are nonetheless characteristic of correlated disorder,\cite{Welberry_1985,Keen2015} and indeed the reported structural model for Zn(hba) contained both positional disorder of the Zn$^{2+}$ ions and orientational disorder of the hba$^{2-}$ ligands. We present here a new set of X-ray diffraction measurements for this system and use them to develop a Truchet-tile description of Zn(hba). Ultimately, what this description captures is that correlated disorder in Zn(hba) is governed by a combination of one-dimensional order of polar ligands and ice-like rules in Zn$^{2+}$ ion coordination. Monte Carlo simulations allow the generation of structural models of Zn(hba) based on the Truchet tilings, and we show these reproduce the experimental diffuse scattering features. A broader implication of our work is that Truchet tilings may provide a useful structural language for a wider class of disordered solids than previously recognised. In this vein, our findings may encourage reinterpretation of other disordered crystal structures in a new light.

\section{Results and Discussion}
\subsection{Average structure: then and now}

Our starting point is to revisit the average structural description of Zn(hba) before eventually proceeding to describe its local structure and correlated disorder. Because the average structure of Zn(hba) is complex, we begin by considering first the related coordination polymer Li(inox) (inox$^-$ = the \textit{N}-oxide of the isonicotinate anion). This Li$^+$-containing system was actually reported in the same study as that of Zn(hba).\cite{White2015} Its structure is a conventionally-ordered analogue of Zn(hba), which is why it is a convenient reference point for the more complex Zn(hba) on which we will come to focus.

\begin{figure}[t]
    \includegraphics{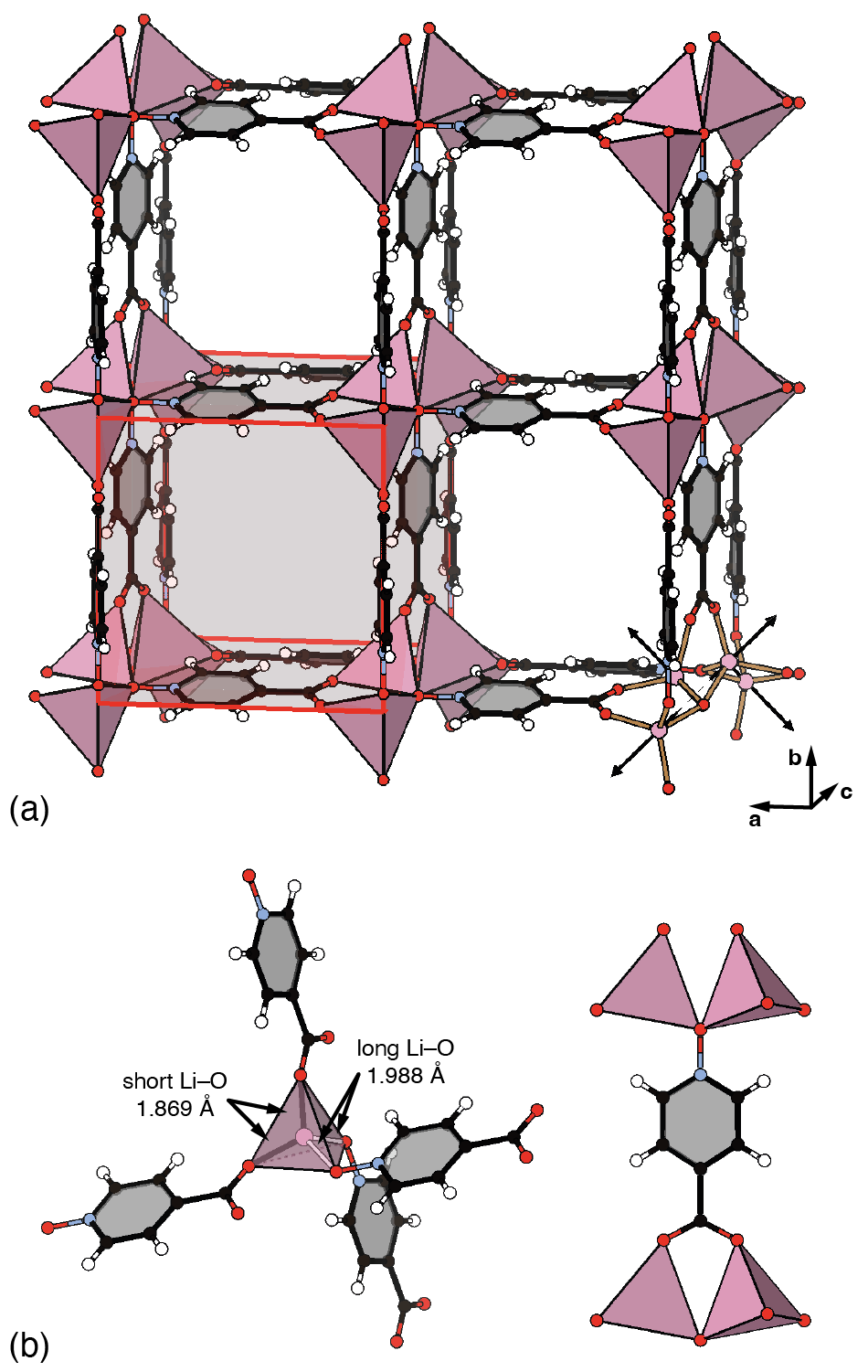}
    \caption{\footnotesize (a) A representation of the crystal structure of Li(inox). Li coordination environments are shown as pink tetrahedra. The C, H, N, and O atoms of the inox$^-$ ligand are shown in black, white, blue, and red, respectively. The unit cell is outlined in red and its interior is shaded. Note the Li centres are displaced away from the cell edges along local $\langle110\rangle$ directions (bottom right). (b) The local coordination of the Li$^+$ (left) and inox$^-$ (right) ions in Li(inox).}
    \label{fig2}
\end{figure}

The crystal structure of Li(inox) is characterised by both long-range dipolar ordering of the inox$^-$ ligands and a two-short-two-long tetrahedral coordination of the Li$^+$ ions. The system crystallises in the chiral tetragonal space group $P4_122$ (or its enantiomorph, $P4_322$) as an ordered {\bf pts} network ($a\sim9$\,\AA, $c\sim12$\,\AA). Along the unique ($\mathbf c$) axis, the structure contains one-dimensional channels with square apertures [Fig.~\ref{fig2}(a)]. These channels are surrounded by perpendicular rows of inox$^-$ chains running along the $\langle100\rangle$ directions. Each Li$^+$ ion, located near a channel vertex, is tetrahedrally coordinated by four inox$^-$ ligands. This coordination environment involves two short and two long Li--O bonds that correspond, respectively, to carboxylate ($d_{\textrm{Li--O}}$ = 1.869(4)\,\AA) and \textit{N}-phenoxide ($d_{\textrm{Li--O}}$ = 1.988(4)\,\AA) donors [Fig.~\ref{fig2}(b)]. In turn, each inox$^-$ ligand binds four Li$^+$ ions in a coplanar arrangement, with two bound at each of its carboxylate and \textit{N}-phenoxide ends [Fig.~\ref{fig2}(b)]. The presence of two short and two long coordination bonds in Li(inox) is enforced by ligand ordering along the crystallographic $\mathbf a$ and $\mathbf b$ directions. Along these directions, each chain of polar inox$^-$ ligands and coordinated Li$^+$ ions---that is, \{$\cdots$inox--Li$_2$--inox--Li$_2\cdots$\}$_\infty$---is ordered such that the ligands adopt the same orientation. This orientation alternates layer-to-layer, presumably to optimise dipolar interactions. The existence of two orthogonal ligand-chain polarities at each Li$^+$ site leads to a displacement of the Li$^+$ ions away from channel vertices along one of the four diagonal $\langle110\rangle$ directions aligned with the vector sum of these polarities. This displacement direction rotates on ascending any given column of Li$^+$ ions as the ligand orientations alternate, giving rise to the helical arrangement captured by the crystallographic $4_1$ (or $4_3$) screw axes [Fig.~\ref{fig2}(a)].

The crystal structure of Zn(hba) reported previously\cite{White2015} involved a similar structural model with Zn$^{2+}$ replacing Li$^+$ and hba$^{2-}$ replacing inox$^-$. The key difference between the two structure solutions was the need to include additional, partially-occupied, cation and anion sites to model significant residual electron density in the Zn(hba) case. Two observations suggested that the original model merits reinterpretation: (i) the presence of additional ``very broad, streaky'' reflections between the Bragg reflections; and (ii) the observation of additional systematic absences ($hhl$, $l$ odd) indicative of an underlying parent structure of higher symmetry.\cite{White2015}

\begin{figure}[t]
    \includegraphics{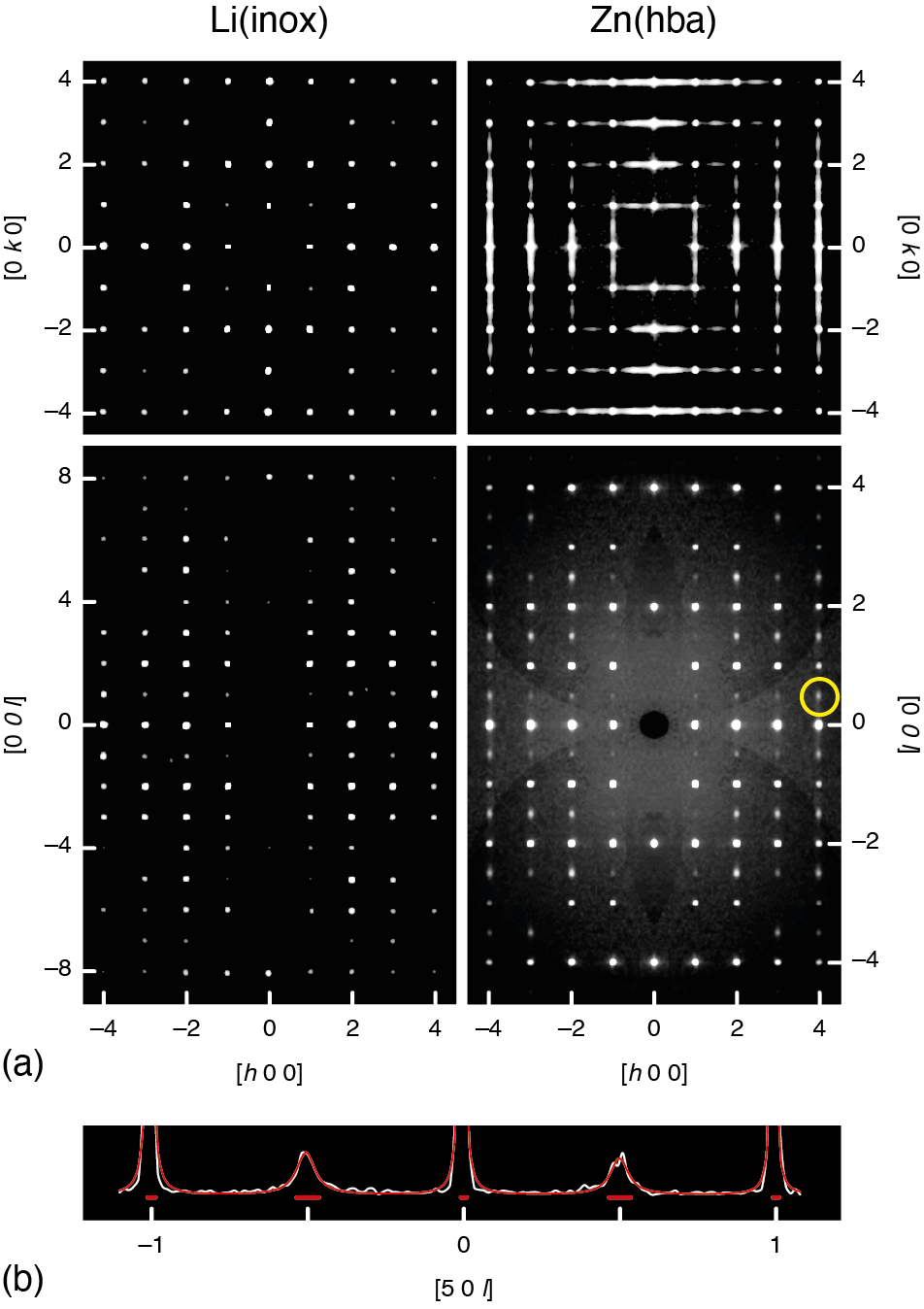}
    \caption{\footnotesize (a) $(hk0)$ (top) and $(h0l)$ (bottom) sections of the single-crystal X-ray diffraction patterns of Li(inox) (left) and Zn(hba) (right). Structured diffuse scattering is evident in the latter but not in the former. This diffuse scattering is transverse polarised in the $(hk0)$ plane; \emph{i.e.} vertical streaks are strongest in a horizontal direction, and \emph{vice versa}. In the $(h0l)$ plane, the diffuse scattering appears as weak Bragg-like features in between the Bragg reflections; an example is circled in yellow. (b) A rocking curve taken along the $[50l]$ direction (white line) is shown fitted using Lorentzian peak shapes (red line). The full-width-at-half-maximum values extracted from these fits are shown as horizontal red bars, and distinguish the Bragg reflections at integral $l$ from the broader diffuse features at $l=\pm\frac{1}{2}$.}
    \label{fig3}
\end{figure}

Hence we sought to reinterpret the structure of Zn(hba) by growing X-ray quality crystals, measuring the corresponding X-ray diffraction pattern, and then redetermining the average structure. Our data for both Li(inox) and Zn(hba) are shown in Fig.~\ref{fig3}. We found that integrating our Zn(hba) data according to the $P4_122$ unit cell used in Ref.~\citenum{White2015} led us to precisely the same interpretation described above. However, our newly acquired data allowed us to distinguish more clearly the Bragg and diffuse scattering features. We found two main contributions to the diffuse scattering. The first and most obvious appeared in the $(hk0)$ plane as transverse-polarised streaks intersecting the $[h00]$ and $[0k0]$ reflections. The second appeared in orthogonal planes (e.g., ($h0l$)) as diffuse scattering maxima at odd values of $l$ that were hitherto interpreted as Bragg reflections, but are in fact diffuse features characteristic of short-range structural correlations. The distinction between Bragg ($l = 2n$) and diffuse ($l = 2n + 1$) scattering in this part of the diffraction pattern can be seen in the corresponding rocking curves [Fig.~\ref{fig3}(b)]. Hence the true average structure of Zn(hba) is that derived by integrating only the Bragg scattering intensity, with the diffuse features sensitive to modulations of this structure with short-range correlations.

\begin{figure}[b]
    \includegraphics{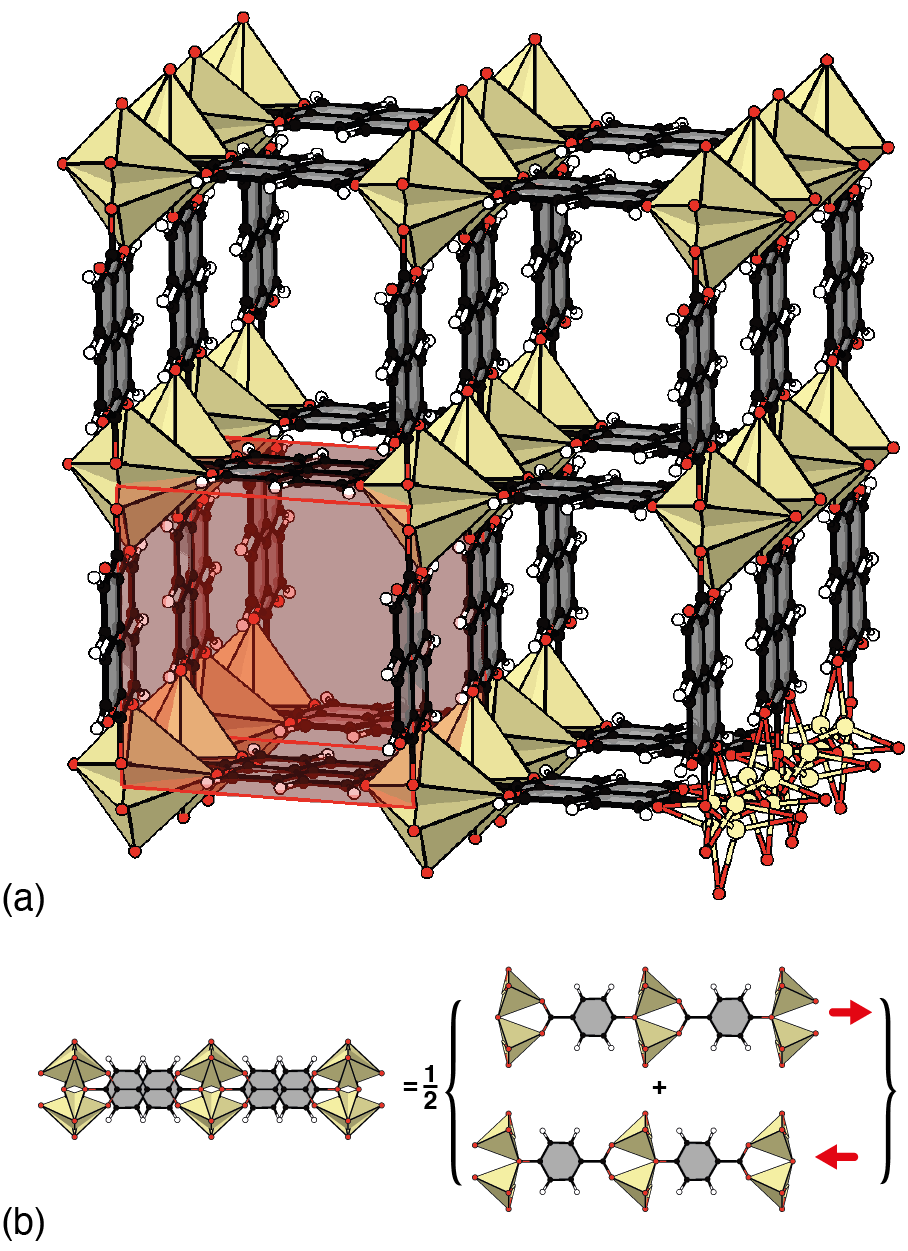}
    \caption{\footnotesize (a) Representation of the average crystal structure of Zn(hba), viewed slightly away from the $\mathbf c$ axis. Zn coordination environments are shown as yellow polyhedra. The C, H, and O atoms of the hba$^{2-}$ ligand are shown in black, white, and red, respectively. The unit cell is outlined in red and its interior is shaded. Note that the Zn sites are now disordered over four positions, each displaced away from the cell edges along one of the $\langle110\rangle$ directions (bottom right). The linker site is also disordered, and contains two inverted orientations of the hba$^{2-}$ ligand. (b) The $\cdots$Zn$_2$-hba-Zn$_2$-hba-Zn$_2\cdots$ chains that run along $\mathbf a$ and $\mathbf b$ correspond to the superposition of two opposite polarisation states. In order to preserve sensible Zn$^{2+}$ coordination geometries, the orientation of each hba ligand within a chain must be the same.}
    \label{fig4}
\end{figure}

Viewed through this new lens, the diffraction patterns revealed much stricter Bragg reflection conditions than previously recognised, reflecting a much higher symmetry average structure solution. The new unit cell is halved along the channel axis direction (i.e., $c\sim6$\,\AA) and has $P4_2/mmc$ crystal symmetry [Fig.~\ref{fig4}(a)]. The existence of a $c$-glide perpendicular to the $\langle110\rangle$ directions accounts for the unexplained systematic absences in the $(hhl)$ planes noted above. Our revised model contains a considerably simpler asymmetric unit: it features one Zn$^{2+}$ ion with 25\% occupancy and half an hba$^{2-}$ ligand with 50\% occupancy. As in the Li(inox) structure, the Zn$^{2+}$ ions in Zn(hba) are displaced away from the network vertices along $\langle110\rangle$ directions. However, these displacements are now disordered such that the parent structure contains four symmetry-equivalent Zn$^{2+}$ positions of equal occupancy. Likewise, the hba$^{2-}$ ligand is orientationally disordered, with the two possible polarisation states occurring with equal occupancy.

Our expectation is that the disordered Zn(hba) structure contains the same local bonding rules as observed in Li(inox). Indeed, the newly-determined average structure model is consistent with each Zn$^{2+}$ ion forming two short and two long Zn--O bonds, corresponding to the carboxylate and phenoxide donors, respectively. Satisfyingly, this simple chemical constraint forces one-dimensional ordering along each ligand chain. It is not possible to reverse the ligand polarities within a \{$\cdots$hba--Zn$_2$--hba--Zn$_2\cdots$\}$_\infty$ chain without either breaking the two-short-two-long rules or heavily distorting the Zn$^{2+}$ coordination geometry [Fig~\ref{fig4}(b)]. Accordingly, from crystal chemical considerations alone, we expect the disorder present in Zn(hba) to be correlated and not random, which is consistent with the observation of structured diffuse scattering in our data. Our challenge now is to show how such a correlated-disordered model of Zn(hba) might be constructed in practice and to show consistency between any such model and the observed scattering patterns.

\subsection{Voronoi decomposition and Truchet tiling}

\begin{figure}[t]
    \includegraphics{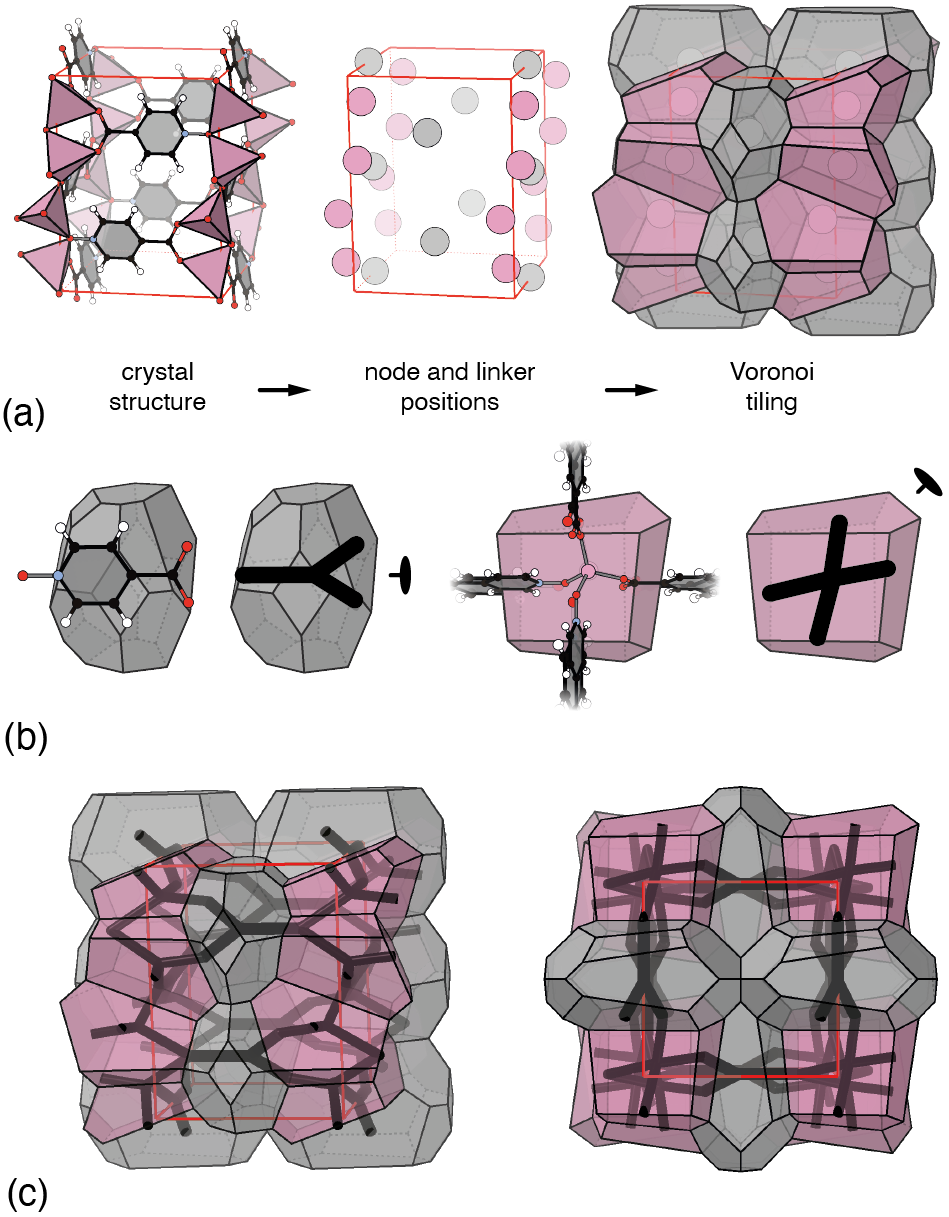}
    \caption{\footnotesize (a) The process of assigning to a MOF structure a corresponding tiling involves first identifying the (average) node and linker positions and then calculating the corresponding Voronoi decomposition. Note that, by design, the Voronoi tiles so obtained inherit the same point symmetries as those of the node and linker positions. The decomposition for Li(inox) shown here produces low symmetry node and linker tiles (shown in pink and grey, respectively) that are programmed to self-assemble into an arrangement related to the Li(inox) structure. (b) We decorate the interior of these tiles according to the chemical connectivity. Because Li(inox) is ordered, these decorations do not break the symmetry of the tiles in this case. Both linker and  node tiles have a single diad that points along the polar axis in the former case and relates the two short and two long bonds in the latter case. (c) Two representations of the decorated-tile description of the Li(inox) structure. The patterns so produced are periodic, and hence are not Truchet tilings.}
    \label{fig5}
\end{figure}

To construct such a model, we interpret the structure of Zn(hba) in terms of a Truchet tiling. We first carry out a Voronoi decomposition of the average structure to generate a set of node and linker tiles, and subsequently consider how the structural elements of Zn(hba) decorate these tiles. By way of background, a Voronoi decomposition takes a set of points and generates a corresponding tiling that fills space.\cite{Aurenhammer1991} In three dimensions, each Voronoi tile is a convex polyhedron that represents the volume of space nearest exactly one of the points in the set. Voronoi decompositions of periodic point sets automatically yield periodic tilings with the same space-group symmetry. When interpreting MOF structures in this way, we use the centres of mass of the node and linker positions as the corresponding set of points [Fig.~\ref{fig5}(a)]. The corresponding Voronoi polyhedra then inherit the point symmetry of the crystallographic sites on which the nodes and linkers are arranged. In disordered MOFs, these polyhedral point symmetries will generally be higher than those of the (chemical) nodes and linkers themselves. Hence, decorating the Voronoi tile according to the molecular structure of its corresponding node or linker fragment breaks its symmetry and a Truchet tiling emerges. Because this methodology is not yet widely used in the MOF literature, we take the opportunity here to illustrate its effect on the ordered Li(inox) structure before applying it to Zn(hba).

In the crystal structure of Li(inox), the Li$^+$ and inox$^-$ components occupy the $4c$ and $4b$ Wyckoff positions of its $P4_122$ unit cell. These positions have point symmetry $2$, with the two-fold axis parallel to a $\langle 110\rangle$ direction for the Li$^+$ site and a $\langle100\rangle$ direction for the inox$^-$ site. The corresponding Voronoi tiles inherit these same point symmetries, giving distorted polyhedra that are essentially programmed to generate the helical node arrangements found in the Li(inox) structure itself [Fig.~\ref{fig5}(a)]. Decorating these tiles with the node and linker geometries preserves their symmetries [Fig.~\ref{fig5}(b)]. Note that the two-fold axis of the $4c$ Li$^+$ site relates each of the two short Li--O or two long Li--O bonds one to the other and also corresponds to the direction along which the Li$^+$ is displaced. Likewise, the two-fold axis at the $4b$ linker site is parallel to the linker polarisation direction. Therefore, the Voronoi tiles themselves also encode the two-short-two-long and polar-chain rules on which we originally rationalised the Li(inox) structure [Fig.~\ref{fig5}(c)].

\begin{figure}[b]
    \includegraphics{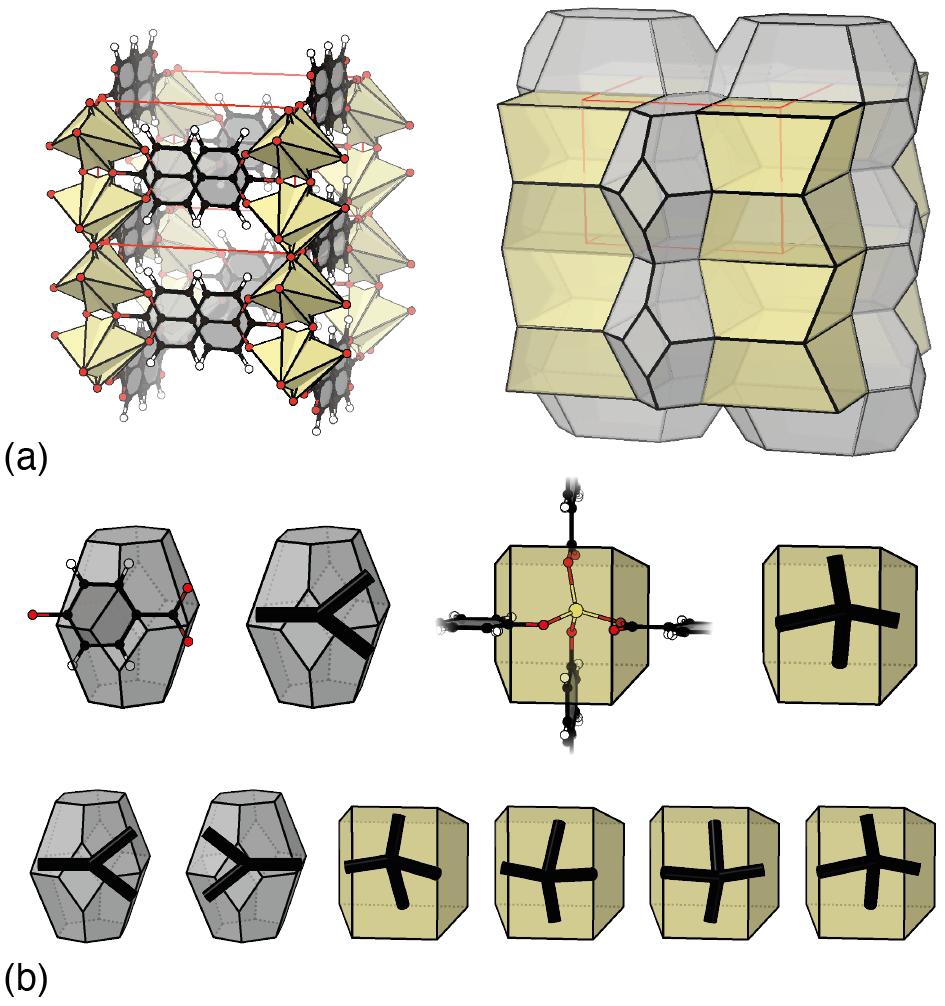}
    \caption{\footnotesize (a) Translation of the Zn(hba) structure into its corresponding Voronoi tesselation generates a higher-symmetry tiling than for Li(inox). Here yellow and grey polyhedra correspond to the node and linker tiles. (b) Decorating these tiles according to the chemical connectivity now lowers the symmetry of the tiles. There are two distinguishable orientations of the linker tile and four of the node tile.}
    \label{fig6}
\end{figure}

Turning now to Zn(hba), we generate its Voronoi tiling using the \emph{average} crystallographic positions of the disordered Zn$^{2+}$ ions and hba$^{2-}$ ligands. The former occupies the $2e$ Wyckoff position (point symmetry $\bar{4}m2$) of the $P4_2/mmc$ cell and the latter the $2c$ position (point symmetry $mmm$). The Voronoi polyhedra so obtained are correspondingly much more symmetric than in the Li(inox) case, yet still generate a packing that is very obviously closely related because their underlying connectivity is still based on the same {\bf pts} net [Fig.~\ref{fig6}(a)]. Decorating these tiles according to the local chemical connectivity now lowers their symmetry [Fig.~\ref{fig6}(b)]. Taking the polyhedron associated with the Zn$^{2+}$ nodes as an example, the existence of two short Zn--O and two long Zn--O bonds breaks both the $\bar{4}$ axis and $\langle 100\rangle$ mirror planes. We use a single one of the four disordered Zn$^{2+}$ positions within this tile to generate the decoration, with the corresponding symmetry breaking most easily seen in the off-centering of the position along a $\langle 110\rangle$ direction. Likewise, at the linker site the polar hba$^{2-}$ ligand generates a connectivity that breaks inversion symmetry of the $mmm$-symmetric Voronoi tile. Here, we use a triangular decoration that represents the acentric connectivity from the hba$^{2-}$ centre-of-mass to the ligand O donor atoms. By breaking the symmetry in this way, there are now four distinguishable orientations of the decorated node tile and two of the decorated linker tile.

\begin{figure}[t]
    \includegraphics{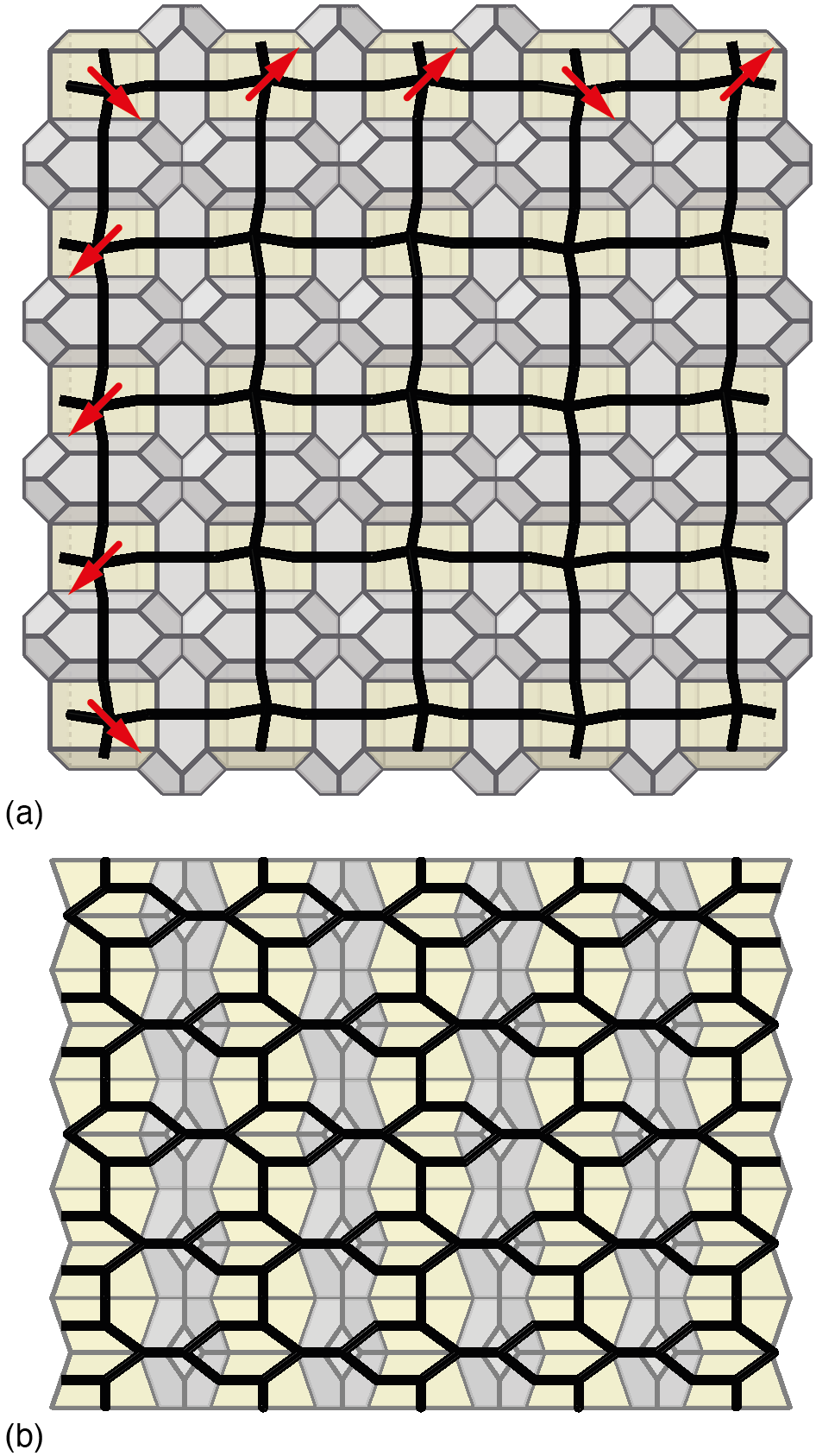}
    \caption{\footnotesize Slices through a representative Truchet-tile arrangement generated for Zn(hba), viewed down (a) $\mathbf c$ and (b) $\mathbf a$. Red arrows show the corresponding Zn displacement directions for the first row and column of node tiles in (a). Note that all arrows in a given row share a common projection along the row axis, as do the arrows in a given column along the column axis.}
    \label{fig6b}
\end{figure}

\begin{figure}[t]
    \includegraphics{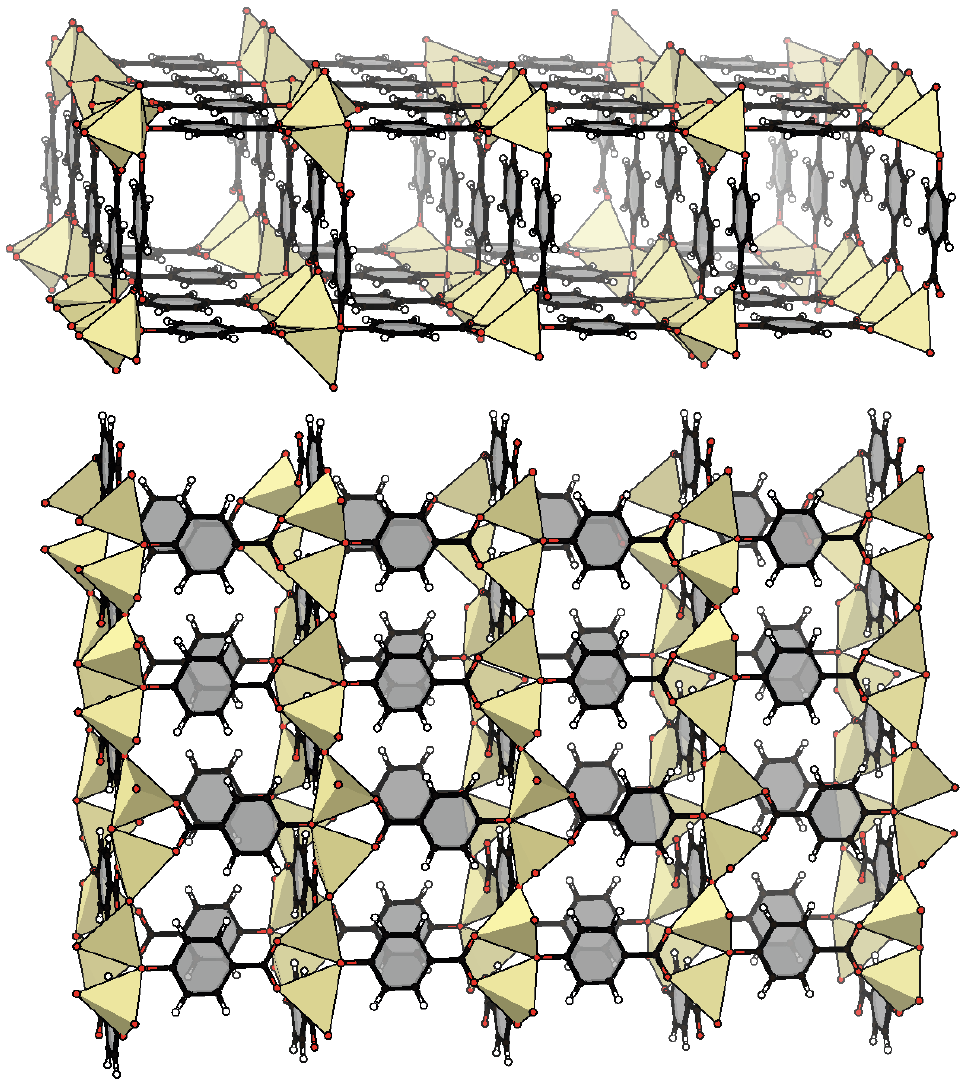}
    \caption{\footnotesize Representations of a fragment of one possible realisation  of the Zn(hba) structure, with connectivity determined by applying the Truchet-tile rules developed in the text. Note the sensible coordination geometries and local connectivities of both Zn$^{2+}$ and hba$^{2-}$ ions.}
    \label{fig7}
\end{figure}

The Truchet-tile description of Zn(hba) then emerges by considering packings of these decorated tiles that obey their natural matching rules [Fig.~\ref{fig6b}]. Packings with randomly-chosen orientations of each tile are unphysical because they inevitably violate these rules. Rather, the tile orientations must be correlated---at least in some directions. For example, the orientation of a Zn$^{2+}$-node tile dictates the dipole direction of the two orthogonal rows of hba-linker tiles in which it lies. Neighbouring rows are free to reorient, however; hence, the matching rules do not enforce long-range periodicity. We show in Fig.~\ref{fig7} a representation of one possible fragment of the Truchet-tile structure of Zn(hba) generated in this way. Note the chemically-sensible connectivity that locally resembles that observed in the ordered structure of Li(inox). The crucial test of this interpretation is whether such models can account for experiment.

\subsection{Monte Carlo simulations and diffuse scattering calculations}

There are two key aspects of the diffuse scattering that will inform our model building approach. Considering first the ($hk0$) plane, we note that the form of the diffuse scattering in this plane consists of lumpy rods running along the $\langle100\rangle^\ast$ directions. The one-dimensional nature of these rods is consistent with strong one-dimensional correlations along the $\langle100\rangle$ directions in the structure itself (i.e., as expected for the hba$^{2-}$ ligand rows), with only weak correlations between these directions. The ``lumpiness'' in this scattering has its maxima at the midpoint between neighbouring reflections, which indicates that there is a slight tendency for neighbouring hba ligand rows to alternate their direction along $\mathbf a$ or $\mathbf b$. The predominant origin of this diffuse scattering is displacive: we can tell this from its transverse polarisation and increasing strength at higher scattering angles.\cite{Welberry_2022} The second observation we make comes from the ($h0l$) plane, where the diffuse scattering appears as Bragg-like features at $l = (2n + 1) / 2$ positions; these are the features originally interpreted as Bragg reflections ($l$ odd) in the context of a doubled-$c$ solution. Again, the presence of these features halfway between Bragg reflections implies a tendency to invert hba ligand row orientations also along the $\mathbf c$ axis. This tendency must be very much stronger than that within the $(a,b)$ plane in order to generate diffuse features that appear almost Bragg-like. Indeed, their width implies a finite correlation length of $ca$~50\,\AA. In the supporting information, we show how the 3D-$\Delta$PDF formalism\cite{Weber_2012}---which involves direct Fourier transform of the diffuse scattering---provides direct visualisation of the structural correlations we describe here.

We proceeded to use a two-step Monte Carlo (MC) approach to translate these observations into representative atomistic models of Zn(hba) for validation against experiment. The first step involved assigning a set of hba ligand row polarisations that reflects the inferences from the diffuse scattering. Using simulation boxes $ca$~50\,\AA~in each direction, we treated the row polarisations within a single $(a,b)$-layer as the MC degree of freedom. Polarisation flips were proposed and accepted or rejected according to a Metropolis algorithm,\cite{Metropolis1953} with the MC energy given by
\begin{equation}
E_1=J\sum_{\langle i,j\rangle}p_ip_j.\label{hamil}
\end{equation}
Here, $p_i=\pm1$ is the polarisation of row $i$ and the sum is taken over neighbouring rows. Carrying out these simulations at an effective Monte Carlo temperature $T_{\rm MC}=2J$ gave row polarisations with weak antipolar order (note the coupling term $J$ was positive). This temperature places the system well within the disordered regime while retaining the weak antipolar correlations implied by experiment. Having assigned polarisations within a single $(a,b)$ layer, we propagated these polarisations along $\mathbf c$, reversing from one plane to the next. Using an odd number of layers introduced a single stacking fault and hence allowed us to simulate the effect of the $\sim50$\,\AA\ correlation length identified above.

Our final task was to translate these polarisation states into an atomistic model that includes their effect on atom displacements. We decorated the polarisation-state configurations with hba coordinates and assigned Zn$^{2+}$ positions on the basis of their neighbouring hba orientations. Note that the use of row polarisations as collective variables in the first step automatically enforces the two-short-two-long coordination rule at each Zn$^{2+}$ centre. The second phase of MC simulations then involved assigning harmonic springs to each Zn--O connection, giving a new MC energy term
\begin{equation}
E_2=\frac{1}{2}k\sum_{\textrm{bonds}}(r-r_{\rm e})^2.\label{springs}
\end{equation}
Here, $r$ is the observed Zn--O bond length and $r_{\rm e}$ the ideal value ($\sim 1.95$\,\AA\ for carboxylate donors and $\sim 2.01$\,\AA\ for phenoxide donors). The effective force constant $k$ acts to scale the magnitude of allowed displacements. We found a value of about 3\,eV\,\AA$^{-2}$ gave physically sensible displacements in simulations at a MC temperature of 300\,K. Our results do not depend on the specific value of this constant.

\begin{figure}[t]
    \includegraphics{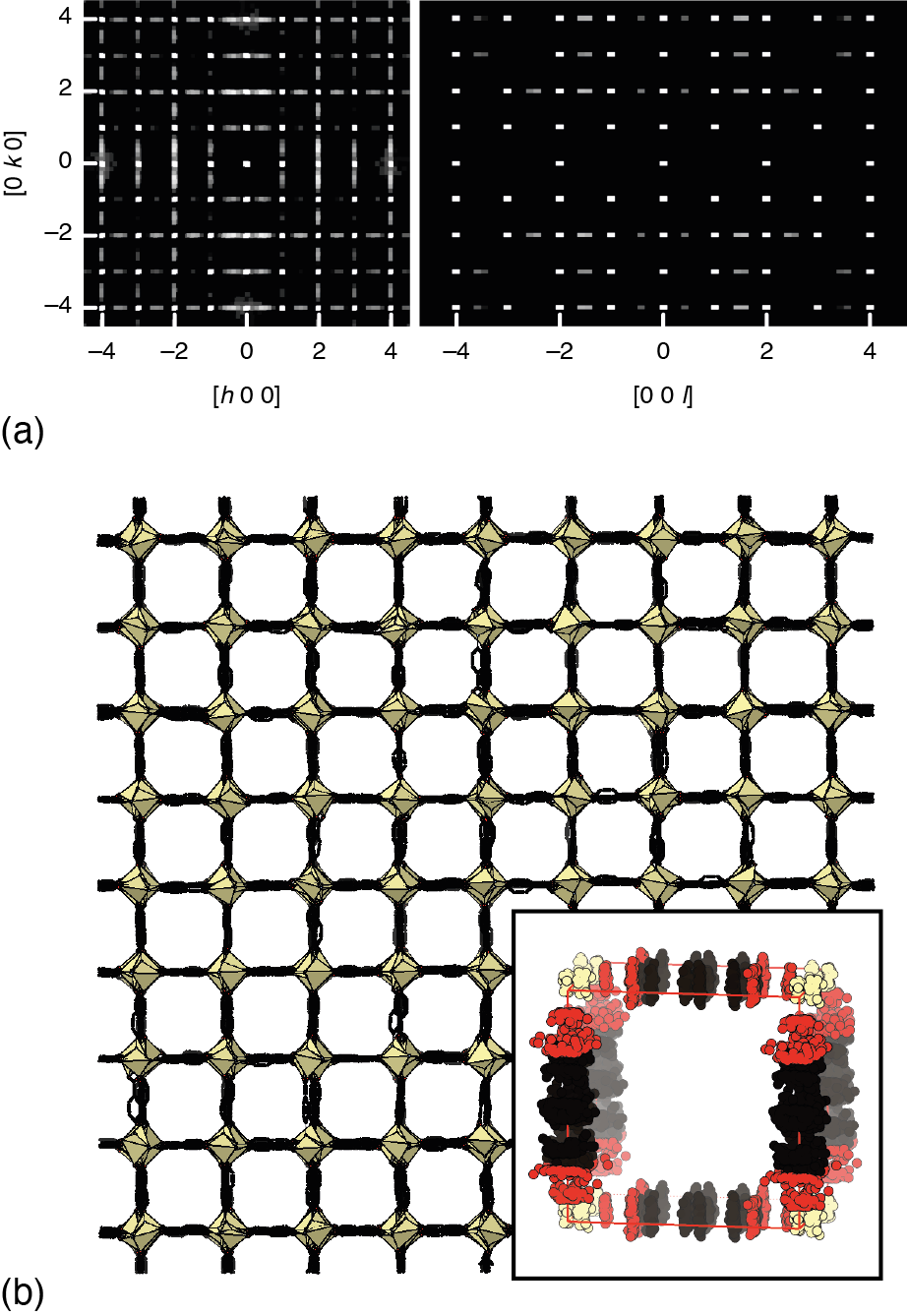}
    \caption{\footnotesize (a) Single-crystal X-ray diffraction patterns calculated from the Truchet-tile Zn(hba) configurations relaxed using direct Monte Carlo simulation. Note the presence of transverse-polarised diffuse scattering in the $(hk0)$ plane and broad maxima at $l=(2n+1)/2$ in the $(0kl)$ plane, as observed experimentally. (b) A representative snapshot of one of the Monte Carlo configurations used to calculate the scattering pattern in (a). The inset shows the projection of this configuration onto a single unit cell. The $P4_2/mmc$ average-structure symmetry determined experimentally emerges naturally despite the presence of disorder.}
    \label{fig8}
\end{figure}

The X-ray diffuse scattering calculated from ensembles of configurations generated in this way is shown in Fig.~\ref{fig8}(a). The calculations reproduce all key experimental features, including both the structured diffuse rods in the ($hk0$) plane and the near-Bragg diffuse maxima observed in ($h0l$) sections. Even greater similarity to experiment might be expected were we able to optimise configurations using density functional theory calculations, which are prohibitively expensive for configurations of the size used here. In Fig.~\ref{fig8}(b), we also show a representative MC configuration together with the corresponding average structure obtained by collapsing the MC supercell onto a single crystallographic unit cell. The local configurations preserve the chemically sensible bonding motifs identified above, whereas the collapsed structure recovers the experimentally-determined average structure. Hence the same model accounts simultaneously for the local chemistry, the crystallographic average structure, and the observed diffuse scattering.

\section{Conclusions}

\begin{figure}[b]
    \includegraphics{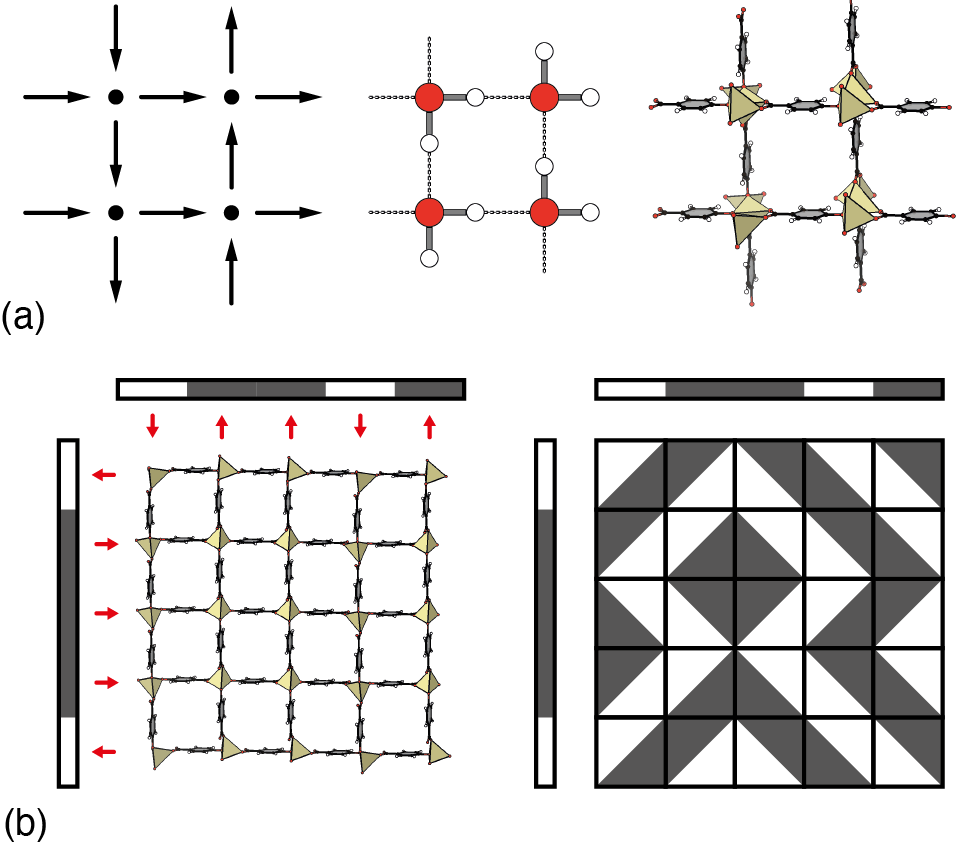}
    \caption{\footnotesize (a) The square ice model developed in Ref.~\citenum{Nagle_1966} consists of a square array of nodes connected by vectors that obey the two-in-two-out constraint derived from hydrogen-bonding rules in water ice. The chain polarisations in Zn(hba) obey the same rules. (b) The ligand orientations and Zn$^{2+}$ ion displacement directions in Zn(hba) are determined by collective chain polarisations. These carry the same information as perpendicular ``barcodes'' associated with the crystal edges. The same barcodes can be used to generate an alternative Truchet-tile representation that encodes the same information.}
    \label{fig9}
\end{figure}

The Truchet tiling rules that we now know to govern correlated disorder in Zn(hba) are actually related to a variant of the long-established ``square-ice'' model studied independently in statistical mechanics [Fig.~\ref{fig9}(a)].\cite{Nagle_1966,Lieb_1967,Lieb_1967b} That particular model concerns a two-dimensional array of nodes arranged on a square lattice, where each node is connected to its four nearest neighbours. Each connection is assigned an Ising degree of freedom that is often interpreted as the orientation of a vector pointing towards one or other of the two nodes it connects. Complexity within the model then emerges by demanding a two-in-two-out rule at each node, whereby exactly two neighbouring edge vectors point in, and the other two point out. The mapping to (water) ice comes by considering a fictitious two-dimensional analogue of cubic ice: O atoms act as nodes of a square lattice and O--H$\cdots$O hydrogen bonds decorate the connections between nodes. The natural bonding constraint that each O is covalently bound to two neighbouring H atoms and hydrogen bonded to two other H atoms leads to the same effective rules realised in Zn(hba).

A particular characteristic of this model is that its configurational entropy is subextensive: the number of collective degrees of freedom scales not with the number of nodes, but rather with its square-root.\cite{Camp_2012} The same is therefore true in Zn(hba), where the arrangement of hba$^{2-}$ ligand polarisations in any two-dimensional sheet, that is, within the $(a,b)$ plane, is uniquely determined by the ligand orientations at the sheet edges.\cite{Overy_2015} Hence the information contained within a crystal of Zn(hba) can be considered a superposition of binary ``barcodes'' running along the $\mathbf a$  and $\mathbf b$ axes [Fig.~\ref{fig9}(b)]. Systems with subextensive entropies are inherently prone to strong size effects---the canonical example being the variation in order/disorder ferroelectric transitions in BaTiO$_3$ samples of different crystal dimensions.\cite{Lambert_1969,Comes_1970,Chaves_1976} In the case of Zn(hba), correlated chain-polarisation disorder will make a greater (relative) contribution to configurational entropy (and hence free energy) in the case of small crystals than in large crystals. We therefore expect different degrees of disorder---and quantitative differences in diffuse scattering---in Zn(hba) crystals of different sizes. We have not yet explored this point experimentally.

There are also good reasons why disorder in Zn(hba) might be controlled chemically. For example, one consequence of the correlated disorder in Zn(hba) is the existence of four distinct types of pore channels, corresponding to different relative orientations of the hba$^{2-}$ ligands surrounding each pore. Considering the crystallographic site at the centre of the channel ($2f$ Wyckoff position), these different pore types reduce the local point symmetry from $\bar{4}m2$ to $\bar{4}$, $222$, $2$, or $1$. Hence, host--guest interactions in Zn(hba) will differ amongst these cases, potentially allowing for guest-driven biasing towards particular ordered (or disordered) daughter configurations during synthesis---for example, as seen in DUT-8.\cite{Ehrling2021}.

\begin{figure}[t]
    \includegraphics{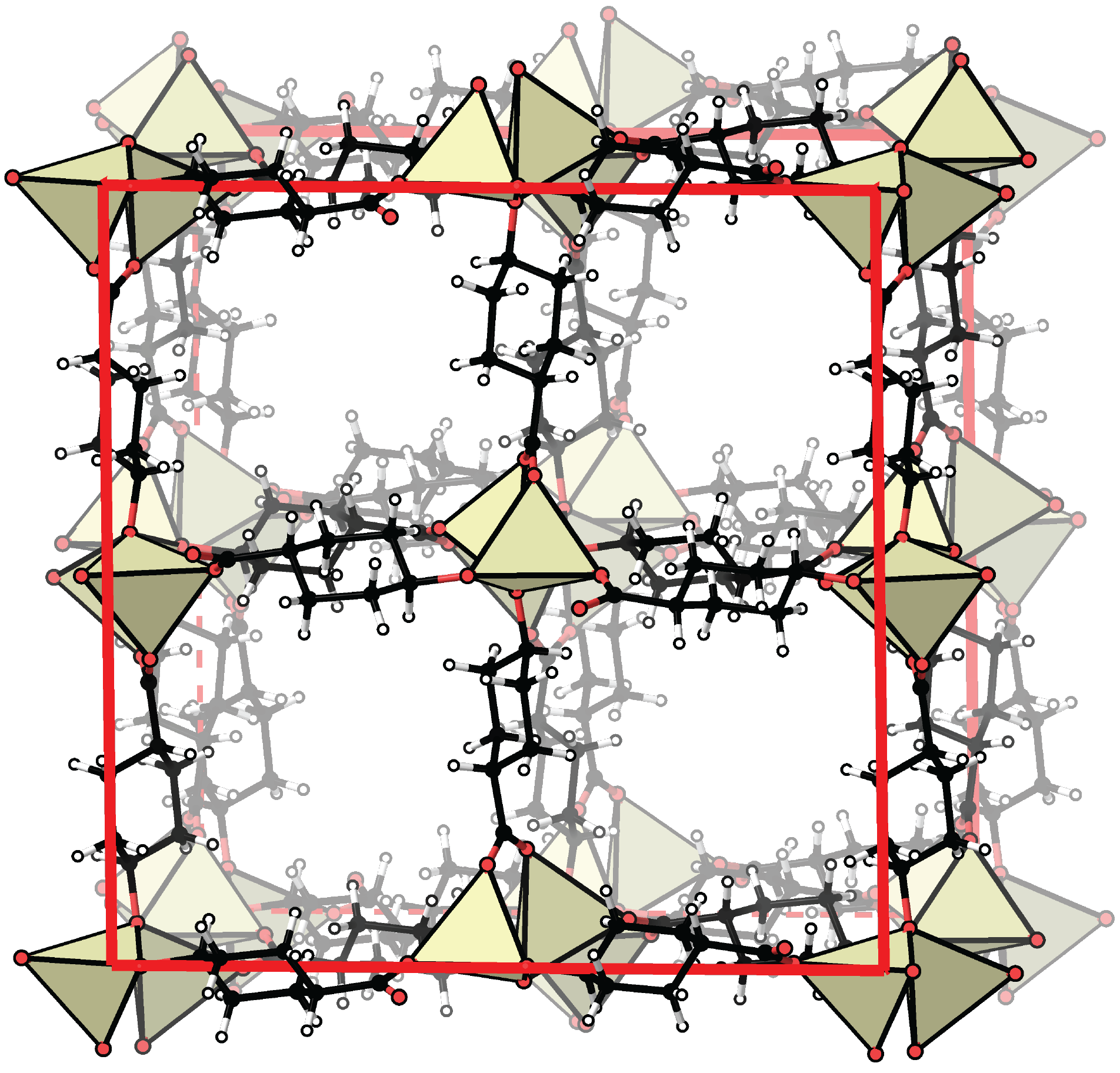}
    \caption{\footnotesize Representation of the crystal structure of Zn(hca), viewed in perspective away from the $\mathbf c$ axis.}
    \label{fig10}
\end{figure}

The observation of long-range order in Li(inox) suggests that compositional control over disorder might also be possible within the Zn-containing analogues. Indeed, we found that replacing hba$^{2-}$ with the related ligand hca$^{2-}$ (hca$^{2-}$ = the dianion of \emph{trans}-4-hydroxycyclohexanecarboxylic acid) gave crystals of Zn(hca) [Fig.~\ref{fig10}], the structure of which was ordered and characterised by the polar tetragonal space group $I4_1cd$ (see SI for further discussion). So, for reasons that are not necessarily obvious, the aromatic hba$^{2-}$ ligand favours disordered configurations while the aliphatic hca$^{2-}$ ligand drives order. Hence, different ligand identities or even combinations thereof seem to allow navigation of the configurational landscape accessible to the Zn(hba) structure type. We note also the existence of disordered analogues obtainable by substituting Co for Zn.\cite{EcheniqueErrandonea_2023} Taken together the existence of these variants allows for precisely the kind of control ultimately required if Truchet architectures are to be exploited for information-storage applications.

Thinking beyond the Zn(hba) family itself, a key result of our study is to demonstrate how the Truchet-tile description initially applied to TRUMOF-1 might serve as a more general paradigm for understanding and interpreting correlated disorder in network structures. Within the MOF field, the process of coarse-graining framework structures into node and linker sites is a mature concept.\cite{Yaghi_2003,Blatov_2007,DelgadoFriedrichs_2007} Hence, the Voronoi decomposition that generates node and linker tiles is a well-defined and deterministic process that might readily be automated. The decorations that add the ``Truchet'' (i.e., symmetry lowering) component do generally rely on chemical intuition or additional knowledge regarding local structure. In the present study, we have used the availability of an ordered analogue (\emph{viz.}~Li(inox)) to infer the local coordination within Zn(hba). The previously mentioned TRUMOF-1 study made use of the obvious compositional similarity to MOF-5.\cite{Meekel2023} In other cases, experimental probes such as NMR, total scattering and/or EXAFS measurements may be crucial to determine the appropriate local structure description.\cite{Meekel2021,Cliffe_2014} One way or the other, we anticipate that Truchet-tile architectures are considerably more widespread than currently recognised. We hope that the approaches developed and applied here will provide a general paradigm through which correlated disorder in crystalline networks may be understood, controlled, and applied---not least in terms of information storage, as originally envisaged by Truchet for the tiles that now bear his name.

\section{Methods}

\textbf{Synthesis.} Reagents and materials were purchased commercially and used as received without further purification. We synthesised Li(inox) according to the method reported in Ref.~\citenum{White2015}. We synthesised Zn(hba) \textit{via} a modified method.

\textit{Li(inox)}. Isonicotinic acid \textit{N}-oxide (0.220\,g, 1.58\,mmol) and LiOH$\cdot$\ce{H2O} (0.070\,g, 1.67\,mmol) were dissolved in boiling MeOH (20\,mL). DMF (20\,mL) was added to the hot reaction mixture. Continued heating at a temperature of 70\,$^\circ$C allowed the selective evaporation of the MeOH from the reaction mixture. The removal of nearly all of the MeOH from the mixture allowed needle-like crystals to separate. The crystals were dried over air before collection.

\textit{Zn(hba)}. Zinc nitrate hexahydrate (16.7\,mg, 0.06\,mmol) and 4-hydroxybenzoic acid (9.1\,mg, 0.07\,mmol) were added to a glass scintillation vial. \textit{N},\textit{N}-Dimethylformamide (4\,mL) was added to the vial with dissolution of the reagents assisted by sonication. The vial was capped and heated at 115\,$^\circ$C for 18--24\,h to yield orange block- or plate-like crystals.

\textit{Zn(hca)}. Zinc nitrate hexahydrate (14.5\,mg, 0.05\,mmol) and \emph{trans}-4-hydroxycyclohexanecarboxylic acid (8.1\,mg, 0.06\,mmol) were added to a glass scintillation vial. \textit{N},\textit{N}-Dimethylformamide (3.5\,mL) and ethanol (0.5\,mL) were added to the vial with dissolution of the reagents assisted by sonication. The vial was capped and heated at 125\,$^\circ$C for 18--24\,h to yield colourless needle-like crystals.

\noindent \textbf{Materials Characterisation.} Single crystal X-ray diffraction data were acquired from a Rigaku XtaLAB Synergy-S Dualflex diffractometer, equipped with a PhotonJet-Cu microfocus source outputting confocal mirror-monochromated Cu-K$\alpha$ radiation ($\lambda= 1.54184$\,\AA), and a DECTRIS EIGER 1M detector. Data were collected about $\omega$ scans in 0.5$^\circ$ increments and were integrated and reduced with absorption corrections being applied through CrysAlisPro (Gaussian grid face-indexed numerical integration with beam profile, or multi-scan).\cite{CRYSALISPRO} Structure solutions were obtained by intrinsic phasing using SHELXT\cite{SHELXT} and were refined by full-matrix least squares on all unique $F^2$ using SHELXL\cite{SHELXL} as implemented within the OLEX2-1.5 GUI.\cite{OLEX2} For Zn(hba), a solvent mask was applied through OLEX2-1.5 to account for the contribution of disordered solvent molecules to the structure
factors. Molecular and crystal structure images were created using CrystalMaker 11 (Version 11.6.2).\cite{CRYSTALMAKER} CCDC deposition numbers: Zn(hba), 2570250; \ce{H2hca}, 2570251; Zn(hca), 2570252.

Room temperature powder X-ray diffraction patterns were acquired from a Bruker D8 Advance Eco diffractometer operating in Bragg--Brentano geometry, outputting Cu-K$\alpha$ ($\lambda= 1.5418$\,\AA) radiation, and equipped with a LYNXEYE XE-T silicon detector. Single crystalline samples of Zn(hba) or Zn(hca) were isolated from their respective crystallisation mother liquors and washed with \textit{N},\textit{N}-dimethylformamide ($3 \times 5$\,mL) to remove any residual ligand or metal salt. The samples were then gently ground to produce polycrystalline powders at which point air-drying at 80\,$^\circ$C was carried out until the point where the majority of surface-bound solvent was removed. Data were acquired over a $2\theta$ range of 5--50$^\circ$.

\noindent \textbf{Voronoi decompositions.} A $3\times3\times3$ supercell configuration of the average node and linker positions in each of Li(inox) and Zn(hba) were generated using CrystalMaker.\cite{CRYSTALMAKER} This supercell size ensured that the central cell was entirely embedded within the configuration. The corresponding node and linker positions were converted from relative to cartesian coordinates (in \AA) using the cell vectors. The resulting coordinates were then used as input for the \emph{voronoin}\cite{Barber1996} function in MATLAB. This routine generates the vertices of Voronoi cells from a given set of input coordinates, but does not support periodic boundary conditions (hence our use of an embedded cell). The resulting Voronoi vertex coordinates were back transformed from cartesian to cell coordinates, and those coordinates contained within the central cell of the supercell identified. These vertex coordinates were reduced to a set of symmetry-distinct coordinates by applying the relevant crystallographic symmetry within CrystalMaker and discarding duplicates.

\noindent \textbf{Monte Carlo Simulations.} We used a two-step Monte Carlo (MC) approach to generate representative atomistic models of Zn(hba). The first step focused on generating a configuration with suitable orientations of hba linkers. The polarisation of each chain (row) of hba ligands within a single $(a,b)$ layer was treated as an Ising variable $p_i = \pm1$. The row polarisations interacted according to Eq.~\eqref{hamil}, where the sum is taken over neighbouring parallel rows and $J > 0$ favours antipolar correlations. Simulations were performed for an $8\times8$ layer with periodic boundary conditions. Trial moves consisted of reversing the polarisation of a randomly-selected row and were accepted or rejected according to the Metropolis criterion. Each simulation was initialised from a random configuration of row polarisations, and subsequently annealed from an effective Monte Carlo temperature $T_{\textrm MC}=20J$ to $2J$. The temperature was multiplied by a factor of 0.95 after every 512 attempted moves. Twenty independent annealing runs were performed.
The final two-dimensional configurations were expanded along the crystallographic $\mathbf c$ direction by reversing all row polarisations between successive layers. Seven layers were used to form a $8\times8\times7$ configuration. The odd number of layers introduced a single stacking fault into the otherwise antipolar stacking sequence, mimicking the $\sim50$\,\AA\ correlation length inferred from the experimental diffuse scattering. The resulting set of linker orientations was then translated into an atomistic model by decorating with hba ligand coordinates, and the Zn$^{2+}$ positions were assigned according to the orientations of the intersecting ligand rows.

In the second step we allowed for thermal displacements. The hba ligands were treated as rigid bodies, and the energy was described by harmonic Zn–O bond potentials as given in Eq.~\eqref{springs}. Equilibrium distances of 1.951 and 2.012\,\AA\ were used for the carboxylate and phenoxide Zn--O bonds, respectively. A common force constant of 3\,eV\,\AA$^{-2}$ was used for both bond types as this was found to produce physically sensible displacements.
Trial moves consisted of either a translation or a rotation. For translations, a randomly selected Zn$^{2+}$ ion or hba ligand rigid body was displaced independently along each Cartesian direction by an amount selected uniformly from the interval $[-1,1]$\,\AA. For rotations, a randomly selected hba ligand was rotated about a randomly oriented axis passing through its centre, with a maximum rotation angle of 5$^\circ$. Periodic boundary conditions were applied, and the unit-cell parameters were held fixed. Moves were accepted or rejected according to the Metropolis criterion. Each simulation consisted of $5\times10^6$ attempted moves. Approximately 20 independent relaxation runs were performed for each of 20 input configurations, and the resulting 347 structures were used for the diffuse-scattering calculations. 

\noindent \textbf{Diffuse Scattering and Calculation of 3D-$\Delta$PDF} Single-crystal diffuse scattering datasets were measured using the same
Rigaku XtaLAB Synergy-S diffractometer and Cu-K$\alpha$ wavelength
described above, collected as a single 360$^\circ$ $\varphi$-scan in
0.2$^\circ$ increments. Three-dimensional reciprocal-space maps were
reconstructed from these data using a custom-written reconstruction
script that relied on the orientation matrix determined in
CrysAlisPro~\cite{CRYSALISPRO}. The reconstructed volume spanned
$\pm 6$ in $h$, $k$, and $l$ on a grid of $601 \times 601 \times 601$
voxels. Subsequent processing was carried out using the program
\textit{meerkat-average}:\cite{simonov-meerkat} the spherically
symmetric background was subtracted and the dataset averaged in the
$4/mmm$ Laue class to improve counting statistics, after which the
Bragg peaks were removed using a punch-and-fill procedure, in which the
intensity within a sphere of 6 voxels centred on each
reciprocal-lattice point was excised and filled with values
interpolated from the surrounding intensity distribution. The resulting
diffuse-only dataset was Fourier transformed to yield the
3D-$\Delta$PDF map~\cite{Weber_2012}. Each signal at a position
$uvw$ in this map corresponds to interatomic pairs separated by the
real-space vector $(u,v,w)$: positive signals identify vectors that
occur more frequently in the real structure than expected from the
average structure, and negative signals identify vectors that occur
less frequently.



\section{Acknowledgements}
This research was supported financially by the UKRI through Frontier Research Grant EP/Z534031/1, by the Royal Society through the Faraday Discovery Fellowships Fund, provided by DSIT, by the Swiss National Science Foundation (Mobility Postdoctoral Fellowship to Y.K.) and by Magdalen College, Oxford (Fellowship by Examination to Y.K.).





\bibliography{references}

@misc{simonov-meerkat,
  author       = {Simonov, Arkadiy},
  title        = {Meerkat: program for reciprocal space reconstruction},
  howpublished = {\url{https://github.com/aglie/meerkat}},
  year         = {2019}
}

@article{EcheniqueErrandonea_2023,
	author = {Echenique-Errandonea, E. and Rojas, S. and Cepeda, J. and Choquesillo-Lazarte, D. and Rodr{\'\i}guez-Di{\'e}guez, A. and Seco, J. M.},
	journal = {Molecules},
	pages = {1846},
	title = {Slow Magnetic Relaxation and Modulated Photoluminescent Emission of Coordination Polymer Based on 3-Amino-4-hydroxybenzoate {Zn} and {Co} Metal Ions},
	volume = {28},
	year = {2023}}

@article{Overy_2015,
	author = {Overy, A. R. and Cairns, A. B. and Cliffe, M. J. and Simonov, A. and Tucker, M. G. and Goodwin, A. L.},
	journal = {Nat. Commun.},
	pages = {10445},
	title = {Design of crystal-like aperiodic solids with selective disorder--phonon coupling},
	volume = {7},
	year = {2015}}

@article{Cliffe_2014,
	author = {Cliffe, M. J. and Wan, W. and Zou, X. and Chater, P. A. and Kleppe, A. K. and Tucker, M. G. and Wilhelm, H. and Funnell, N. P. and Coudert, F.-X. and Goodwin, A. L.},
	journal = {Nat. Commun.},
	pages = {4176},
	title = {Correlated defect nanoregions in a metal--organic framework},
	volume = {5},
	year = {2014}}

@article{DelgadoFriedrichs_2007,
	author = {Delgado-Friedrichs, O. and O'Keeffe, M. and Yaghi, O. M.},
	journal = {Phys. Chem. Chem. Phys.},
	pages = {1035-1043},
	title = {Taxonomy of periodic nets and the design of materials},
	volume = {9},
	year = {2007}}

@article{Blatov_2007,
	author = {Blatov, V. A. and Delgado-Friedrichs, O. and O'Keeffe, M. and Proserpio, D. M.},
	journal = {Acta Cryst. A},
	pages = {418-425},
	title = {Three-periodic nets and tilings: natural tilings for nets},
	volume = {63},
	year = {2007}}

@article{Lambert_1969,
	author = {Lambert, M. and Comes, R.},
	journal = {Solid State Commun.},
	pages = {305-308},
	title = {The chain structure and phase transition of {BaTiO$_3$} and {KNbO$_3$}},
	volume = {7},
	year = {1969}}

@article{Comes_1970,
	author = {Com{\`e}s, R. and Lambert, M. and Guinier, A.},
	journal = {Acta Cryst. A},
	pages = {244-254},
	title = {D{\'e}sordre lin{\'e}aire dans les cristaux (cas du silicium, du quartz, et des p{\'e}rovskites ferro{\'e}lectriques)},
	volume = {26},
	year = {1970}}

@article{Chaves_1976,
	author = {Chaves, A. S. and Barreto, F. C. S. and Nogueira, R. A. and Z{\~e}ks, B.},
	journal = {Phys. Rev. B},
	pages = {207-212},
	title = {Thermodynamics of an eight-site order-disorder model for ferroelectrics},
	volume = {13},
	year = {1976}}

@article{Camp_2012,
	author = {Camp, P. J. and Fuertes, A. and Attfield, J. P.},
	journal = {J. Am. Chem. Soc.},
	pages = {6762-6766},
	title = {Subextensive Entropies and Open Order in Perovskite Oxynitrides},
	volume = {134},
	year = {2012}}

@article{Nagle_1966,
	author = {Nagle, J. F.},
	journal = {J. Math. Phys.},
	pages = {1484-1491},
	title = {Lattice Statistics of Hydrogen Bonded Crystals. {I}. {T}he Residual Entropy of Ice},
	volume = {7},
	year = {1966}}

@article{Lieb_1967,
	author = {Lieb, E. H.},
	journal = {Phys. Rev.},
	pages = {162-172},
	title = {Residual Entropy of Square Ice},
	volume = {162},
	year = {1967}}

@article{Lieb_1967b,
	author = {Lieb, E. H.},
	journal = {Phys. Rev. Lett.},
	pages = {692-694},
	title = {Exact Solution of the Problem of the Entropy of Two-Dimensional Ice},
	volume = {18},
	year = {1967}}

@article{Weber_2012,
	author = {Weber, T. and Simonov, A.},
	journal = {Z. Krist.},
	pages = {238-247},
	title = {The three-dimensional pair distribution function analysis of disordered single crystals: basic concepts},
	volume = {227},
	year = {2012}}

@book{Welberry_2022,
	address = {Oxford},
	author = {Welberry, T. R.},
	edition = {2nd},
	publisher = {Oxford University Press},
	title = {Diffuse {X}-ray scattering and models of disorder},
	year = {2022}}

@article{Welberry_1985,
	author = {Welberry, T. R.},
	journal = {Rep. Prog. Phys.},
	pages = {1543-1593},
	title = {Diffuse x-ray scattering and models of disorder},
	volume = {48},
	year = {1985}}

@article{Robson_2000,
	author = {Robson, R.},
	journal = {J. Chem. Soc., Dalton Trans.},
	pages = {3735-3744},
	title = {A Net-Based Approach to Coordination Polymers},
	year = {2000}}

@article{Yaghi_2003,
	author = {Yaghi, O. M. and O'Keeffe, M. and Ockwig, N. W. and Chae, H. K. and Eddaoudi, M. and Kim, J.},
	journal = {Nature},
	pages = {705-714},
	title = {Reticular synthesis and the design of new materials},
	volume = {423},
	year = {2003}}

@article{Kolmogorov_1968,
	author = {Kolmogorov, A. N.},
	journal = {Int. J. Comp. Math.},
	pages = {157-168},
	title = {Three approaches to the quantitative definition of information},
	volume = {2},
	year = {1968}}

@article{Crutchfield_2011,
	author = {Crutchfield, J. P.},
	journal = {Nat. Phys.},
	pages = {17-24},
	title = {Between order and chaos},
	volume = {8},
	year = {2011}}

@article{Crutchfield_1994,
	author = {Crutchfield, J. P.},
	journal = {Physica D},
	pages = {11-54},
	title = {The calculi of emergence: computation, dynamics and induction},
	volume = {75},
	year = {1994}}

@article{Cartwright_2012,
	author = {Cartwright, J. H. E. and Mackay, A. L.},
	journal = {Phil. Trans. R. Soc. A},
	pages = {2807-2822},
	title = {Beyond crystals: the dialectic of materials and information},
	volume = {370},
	year = {2012}}

@book{Pretzel_1992,
	address = {Oxford},
	author = {Pretzel, O.},
	publisher = {Oxford University Press},
	title = {Error-correcting Codes and Finite Fields},
	year = {1992}}

@article{Gartside_2022,
	author = {Gartside, J. C. and Stenning, K. D. and Vanstone, A. and Holder, H. H. and Arroo, D. M. and Dion, T. and Caravelli, F. and Kurebayashi, H. and Branford, W. R.},
	journal = {Nat. Nano.},
	pages = {460-469},
	title = {Reconfigurable training and reservoir computing in an artificial spin-vortex ice via spin-wave fingerprinting},
	volume = {17},
	year = {2022}}

@article{Keen_2015,
	author = {Keen, D. A. and Goodwin, A. L.},
	journal = {Nature},
	pages = {303-309},
	title = {The crystallography of correlated disorder},
	volume = {521},
	year = {2015}}

@article{Goodwin_2025,
	author = {Goodwin, A. L.},
	journal = {arXiv:},
	pages = {2509.09171},
	title = {Structural Complexity and Correlated Disorder in Materials Chemistry},
	year = {2025}}

@article{Simonov_2020,
	author = {Simonov, A. and Goodwin, A. L.},
	journal = {Nat. Rev. Chem.},
	pages = {657-673},
	title = {Designing disorder into crystalline materials},
	volume = {4},
	year = {2020}}

@article{Griffin_2025,
	author = {Griffin, S. L. and Meekel, E. G. and Bulled, J. M. and Canossa, S. and Wahrhaftig-Lewis, A. and Schmidt, E. M. and Champness, N. R.},
	journal = {Nat. Commun.},
	number = {3209},
	title = {A lanthanide {MOF} with nanostructured node disorder},
	year = {16}}

@article{Meekel2023,
	author = {Emily G. Meekel and Ella M. Schmidt and Lisa J. Cameron and A. David Dharma and Hunter J. Windsor and Samuel G. Duyker and Arianna Minelli and Tom Pope and Giovanni Orazio Lepore and Ben Slater and Cameron J. Kepert and Andrew L. Goodwin},
	journal = {Science},
	pages = {357--361},
	title = {Truchet-tile structure of a topologically aperiodic metal--organic framework},
	volume = {379},
	year = {2023}}

@article{Meekel2021,
	author = {Emily G. Meekel and Andrew L. Goodwin},
	journal = {CrystEngComm},
	pages = {2915--2922},
	title = {Correlated disorder in metal--organic frameworks},
	volume = {23},
	year = {2021}}

@article{Dixey2019,
	author = {R. J. C. Dixey and F. Orlandi and P. Manuel and P. Mukherjee and S. E. Dutton and P. J. Saines},
	journal = {Philos. Trans. R. Soc., A},
	pages = {20190007},
	title = {Emergent magnetic order and correlated disorder in formate metal--organic frameworks},
	volume = {377},
	year = {2019}}

@article{Ehrling2021,
	author = {Sebastian Ehrling and E. M. Reynolds and Volodymyr Bon and Irena Senkovska and Tobias E. Gorelik and Jonathan D. Evans and Martin Rauche and Matthias Mendt and Matthias S. Weiss and Andreas P{\"o}ppl and Erdmann Brunner and Ute Kaiser and Stefan Kaskel and Andrew L. Goodwin},
	journal = {Nat. Chem.},
	pages = {568--574},
	title = {Adaptive response of a metal--organic framework through reversible disorder--disorder transitions},
	volume = {13},
	year = {2021}}

@article{James2003,
	author = {Stuart L. James},
	journal = {Chem. Soc. Rev.},
	pages = {276--288},
	title = {Metal--organic frameworks},
	volume = {32},
	year = {2003}}

@article{Yaghi1998,
	author = {Omar M. Yaghi and Hailian Li and Charles Davis and David Richardson and Thomas L. Groy},
	journal = {Acc. Chem. Res.},
	pages = {474--484},
	title = {Synthetic Strategies, Structure Patterns, and Emerging Properties in the Chemistry of Modular Porous Solids},
	volume = {31},
	year = {1998}}

@article{Truchet1704,
	author = {S. Truchet},
	journal = {M{\'e}m. Acad. R. Sci. (\textit{Paris})},
	pages = {363--372},
	title = {M{\'e}moir sur les combinasions},
	volume = {1704},
	year = {1704}}

@article{White2015,
	author = {Keith F. White and Brendan F. Abrahams and Ravichandar Babarao and A. David Dharma and Timothy A. Hudson and Helen E. Maynard-Casely and Richard Robson},
	journal = {Chem. Eur. J.},
	pages = {18057--18061},
	title = {A New Structural Family of Gas-Sorbing Coordination Polymers Derived from Phenolic Carboxylic Acids},
	volume = {21},
	year = {2015}}

@article{Keen2015,
	author = {David A. Keen and Andrew L. Goodwin},
	journal = {Nature},
	pages = {303--309},
	title = {The crystallography of correlated disorder},
	volume = {521},
	year = {2015}}

@article{Aurenhammer1991,
	author = {Franz Aurenhammer},
	journal = {ACM Comput. Surv.},
	pages = {345--405},
	title = {Voronoi Diagrams---A Survey of a Fundamental Geometric Data Structure},
	volume = {23},
	year = {1991}}

@article{Metropolis1953,
	author = {Metropolis, Nicholas and Rosenbluth, Arianna W. and Rosenbluth, Marshall N. and Teller, Augusta H. and Teller, Edward},
	journal = {J. Chem. Phys.},
	pages = {1087--1092},
	title = {Equation of State Calculations by Fast Computing Machines},
	volume = {21},
	year = {1953}}

@article{SHELXL,
	author = {G. M. Sheldrick},
	journal = {Acta Cryst.},
	pages = {3--8},
	title = {Crystal structure refinement with SHELXL},
	volume = {C\textbf{71}},
	year = {2015\textit{b}}}

@article{SHELXT,
	author = {G. M. Sheldrick},
	journal = {Acta Cryst.},
	pages = {3--8},
	title = {SHELXT -- Integrated space-group and crystal-structure determination},
	volume = {A\textbf{71}},
	year = {2015\textit{a}}}

@article{OLEX2,
	author = {O. V. Dolomanov and L. J. Bourhis and R. J. Gildea and J. A. K. Howard and H. Puschmann},
	journal = {\textit{J. Appl. Cryst.}},
	pages = {339--341},
	title = {OLEX2: a complete structure solution, refinement and analysis program},
	volume = 42,
	year = 2009}

@misc{CRYSTALMAKER,
	author = {D. Palmer},
	howpublished = {CrystalMaker Software, Bicester, Oxfordshire, England. https://www.crystalmaker.com/},
	title = {{\textit{CrystalMaker}}},
	year = {2020}}

@misc{CRYSALISPRO,
	author = {{Rigaku OD}},
	howpublished = {Rigaku Oxford Diffraction Ltd, Yarnton, Oxfordshire, England.},
	title = {{\textit{CrysAlis PRO}}},
	year = {2025}}

@article{Barber1996,
	author = {C. B. Barber and D. P. Dobkin and H. T. Huhdanpaa},
	journal = {ACM Trans. Math. Softw.},
	pages = {469--483},
	title = {The Quickhull Algorithm for Convex Hulls},
	volume = {22},
	year = {1996}}

\end{document}